\documentclass[prd,longbibliography,nofootinbib,twocolumn,preprintnumbers]{revtex4-1} %

\usepackage[T1]{fontenc}
\usepackage[utf8]{inputenc}
\usepackage{relsize}
\usepackage{soul}
\usepackage{calligra}
\usepackage{verbatim}
\usepackage{caption}
\usepackage{adjustbox}
\usepackage[titles]{tocloft} %
\cftsetindents{section}{0em}{2.5em}
\cftsetindents{subsection}{2.5em}{2.5em}

\usepackage{appendix}

\usepackage{graphicx,epsfig}
\usepackage[usenames,dvipsnames,table]{xcolor}
\usepackage{subfig}
\usepackage{float}
\usepackage{placeins}

\usepackage{tabularx}
\usepackage{makecell,multirow}
\usepackage{dcolumn}
\usepackage{enumitem}

\usepackage{amsmath,amsthm,amssymb}
\usepackage{slashed}
\usepackage{mathrsfs}
\usepackage{bm}
\usepackage{leftidx}
\usepackage{commath}

\usepackage{url}
\usepackage{color}
\definecolor{darkblue}{rgb}{0.0,0.0,0.6}
\definecolor{red}{rgb}{0.9, 0,0}
\definecolor{navy}{rgb}{0.05, 0.05,0.8}
\usepackage[colorlinks]{hyperref}
\hypersetup{
     colorlinks   = true,
     citecolor    = darkblue,
     linkcolor    = darkblue,
     urlcolor    = darkblue
}
\graphicspath{{./figures/}}

\newcommand{\tx}[1]{\ensuremath{\textnormal{#1}}}
\newcommand{\eqa}[1]{\begin{align}#1\end{align}}
\newcommand{\subeqa}[2]{\begin{subequations} \la{#1} \eqa{#2} \end{subequations}}

\def\bit{\begin{itemize}}
\def\eit{\end{itemize}}
\def\ben{\begin{enumerate}}
\def\een{\end{enumerate}}
\newcommand{\la}[1]{\label{#1}}
\newcommand{\Eq}[1]{Eq.~\eqref{#1}} 
 
\newcommand{\Eqs}[2]{Eqs.~\eqref{#1} and \eqref{#2}} 
\newcommand{\Sec}[1]{Sec.~\ref{#1}}

\newcommand{\Fig}[1]{Fig.~\ref{#1}}

\newcommand{\Lag}{\mathscr{L}}

\newcommand{\units}[1]{\;\tx{#1}}
\newcommand{\eV}{\units{eV}}

\newcommand{\cm}{\units{cm}}

\newcommand{\micron}{\mu \text{m}}

\newcommand{\be}{\begin{equation}}
\newcommand{\ee}{\end{equation}}
\newcommand{\nl}{\nonumber \\}
\newcommand{\order}[1]{\mathcal{O}{(#1)}}
\newcommand{\jhat}{\hat{j \, }}

\newcommand{\diffp}[2]{\mathop{}\!\mathrm{d^{#1}}{#2}\mathop{}\!}

\newcommand{\grad}{\vec{\nabla}}

\newcounter{questioncount}[section]

\newcommand{\mA}{m_{A^\prime}}
\newcommand{\eps}{\epsilon}
\newcommand{\inv}{\tx{inv}}
\newcommand{\vis}{\tx{vis}}

\newcommand{\vEvis}{\vec{E}_\vis}

\newcommand{\vBvis}{\vec{B}_\vis}
\newcommand{\om}{\omega}

\newcommand{\vjeff}{\vec{j}_\text{eff}}
\newcommand{\p}{{\, \prime}}
\newcommand{\vxp}{\vec{x}^{\, \prime}}

\newcommand{\vAp}{\vec{A}^{\p}}

\usepackage{calligra}
\DeclareMathAlphabet{\mathcalligra}{T1}{calligra}{m}{n}
\DeclareFontShape{T1}{calligra}{m}{n}{<->s*[2.2]callig15}{}

\definecolor{mpl_firebrick}{rgb}{0.6980392156862745, 0.13333333333333333, 0.13333333333333333}
\definecolor{mpl_maroon}{rgb}{0.5019607843137255, 0.0, 0.0}
\definecolor{mpl_Blues8}{rgb}{0.09019607843137256, 0.39294117647058824, 0.6705882352941177}
\definecolor{mpl_8}{rgb}{0.8, 0.8, 0.8}

\begin{document}

\preprint{FERMILAB-PUB-23-073-SQMS-T}

\title{Light Shining Through a Thin Wall: Evanescent Hidden Photon Detection}
\author{Asher Berlin}
\email{aberlin@fnal.gov}
\author{Roni Harnik}
\email{roni@fnal.gov }
\author{Ryan Janish}
\email{rjanish@fnal.gov}
\affiliation{Theory Division, Fermi National Accelerator Laboratory, Batavia, IL 60510, USA}
\affiliation{Superconducting Quantum Materials and Systems Center (SQMS), Fermi National Accelerator Laboratory, Batavia, IL 60510, USA}

\begin{abstract}

A kinetically-mixed hidden photon is sourced as an evanescent mode by electromagnetic fields that oscillate at a frequency smaller than the hidden photon mass. These evanescent modes fall off exponentially with distance, but nevertheless yield detectable signals in a photon regeneration experiment if the electromagnetic barrier is made sufficiently thin. We consider such an experiment using superconducting cavities at $\text{GHz}$ frequencies, proposing various cavity and mode arrangements that enable unique sensitivity to hidden photon masses ranging from $10^{-5} \ \text{eV}$ to $ 10^{-1} \ \text{eV}$.

\end{abstract}

\maketitle


\section{Introduction}

Massive vector fields with feeble couplings to Standard Model (SM) particles may readily exist and present a compelling target for experimental searches.  Such \emph{hidden photons}~\cite{Holdom:1985ag} have been well-studied both in their own right and as a possible component of the dark matter sector (see, e.g., Refs.~\cite{Fabbrichesi:2020wbt,Alexander:2016aln,Battaglieri:2017aum,Antypas:2022asj} and references within).
A simple and natural possibility is that the hidden photon couples to the SM through a kinetic mixing $\eps \ll 1$,
\eqa{
\la{eq:L-kineticmix}
    \Lag = - \frac{1}{4} \,  F^2
     - \frac{1}{4}  \, F^{\prime \, 2}
     + \frac{1}{2} \, \mA^2 A^{\prime \, 2}
     +  \frac{\eps}{2} \, F F^\prime
     -  j  A
     ~,
}
where $A$ is the SM photon field and $F$ its field strength, $A'$ is the hidden photon field with mass $\mA$ and $F'$ its field strength, and $j$ is the electromagnetic (EM) current density. We have suppressed Lorentz indices and absorbed the EM coupling into $j$ for brevity.

Many experiments have directly searched for or indirectly constrained the existence of such a kinetically-mixed hidden photon~\cite{Alexander:2016aln,Battaglieri:2017aum,Caputo:2021eaa,Antypas:2022asj}.
Most of these efforts have involved searching for the effects of hidden photons that are produced \emph{on-shell}, since a detectable signal often requires propagation across a considerable distance compared to an $A^\prime$ Compton wavelength.\footnote{
Exceptions to this are, e.g., limits derived from tests of Coulomb's law~\cite{Williams:1971ms,Abel:2008ai} and atomic spectroscopy~\cite{Jaeckel:2010xx}.}
Experiments of this form include photon regeneration or ``light-shining-through-wall'' (LSW) experiments, such as ALPS~\cite{Ehret:2010mh} and CROWS~\cite{Betz:2013dza}, as well as Dark SRF, which recently set its first limits employing two ultra-high quality superconducting radio frequency (SRF) cavities~\cite{Romanenko:2023irv}.

In the field basis of \Eq{eq:L-kineticmix}, an LSW setup involves driving an EM field $A$ at a frequency $\om$ which in turn produces hidden photons $A^\prime$.
These hidden photons have energy $\om$ and propagate with momentum $k = \sqrt{\om^2 - \mA^2}$ across a barrier opaque to SM photons, after which they source a SM field $A$ in a shielded detection region.
For an off-shell hidden photon $\mA > \om$, the signal is evanescent, i.e., it is suppressed by the propagation distance~$d$ as $e^{-|k| d}$.
In most cases, $d$ corresponds to the separation between the production and detection regions, which is usually chosen to be $d \gtrsim \om^{-1}$.
Thus, in such an arrangement, searching for more massive hidden photons requires operating EM sources at higher frequencies, since this increases the maximum mass for which such particles can propagate on-shell.
This is the strategy employed in current experiments such as ALPS~\cite{Ehret:2010mh} and recently proposed millimeter wavelength setups~\cite{Miyazaki:2022kxl}.

\begin{figure*}[t]
\centering
\includegraphics[width=13cm]{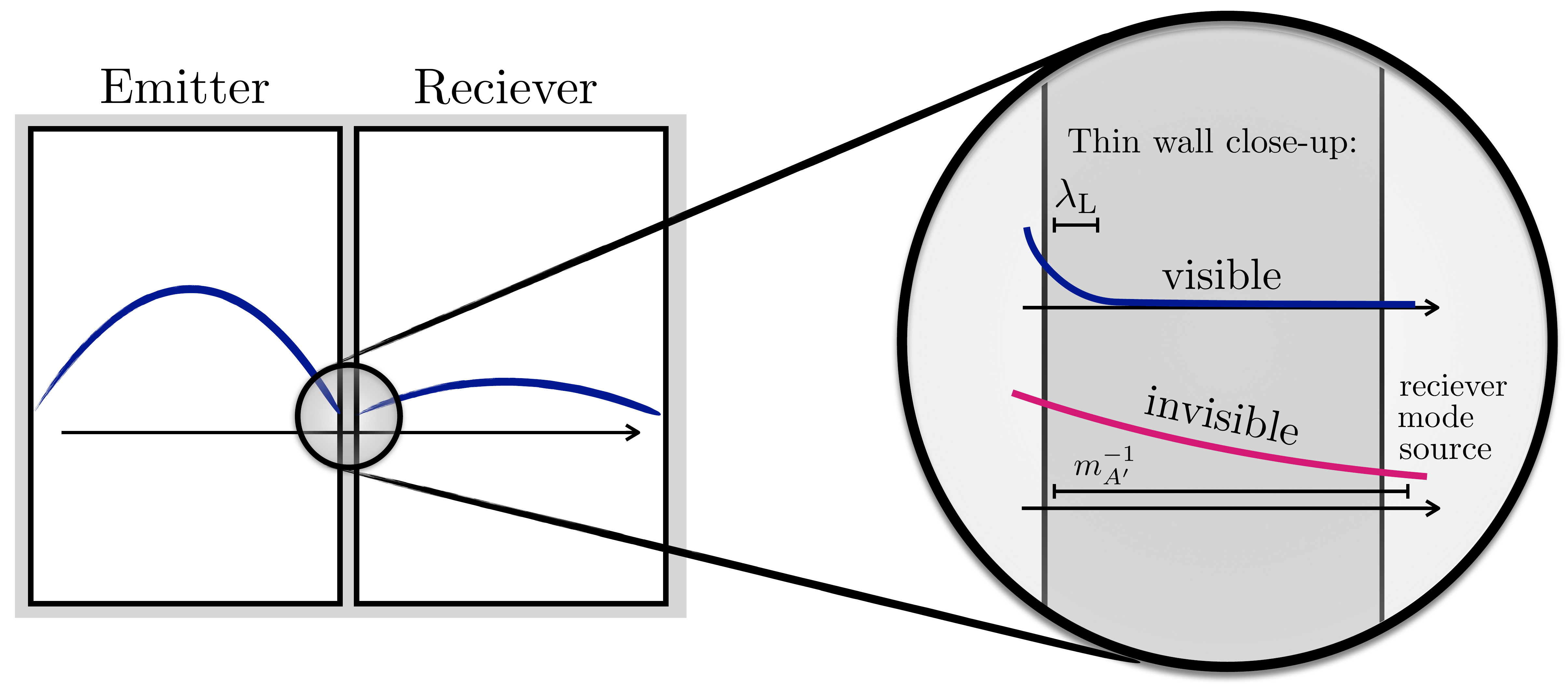}
\caption{\textbf{Left}: A sketch of the LSthinW setup. A highly excited emitter cavity is separated from a quiet receiver cavity by a thin wall. %
The driven emitter and signal receiver modes are shown in blue.
\textbf{Right}: A closeup of the thin wall. The visible photon %
field (blue) is suppressed over the short London penetration depth $\lambda_\tx{L}$ whereas the evanescent invisible mode (pink) extends across the hidden photon Compton wavelength $\sim 1/\mA$, which can be orders of magnitude larger than the wall thickness. The receiver mode is resonantly excited to detectable levels even though it is effectively sourced by the thin region of non-zero invisible field near the receiver side of the wall. }
\label{fig:LSthinWcartoon}
\end{figure*}

However, using higher frequency sources is not a fundamental requirement.
It appears as a consequence of requiring long propagation distances in an LSW experiment, which can be avoided by employing a thin barrier, $d \ll \om^{-1}$.
We are thus motivated to consider light-shining-through-\emph{thin}-wall (LSthinW) setups, which allow for the detection of evanescent (i.e., virtual or off-shell) hidden photon signals for $\mA \gg \om$.
There is considerable advantage in this approach as opposed to increasing the frequency, since lower frequency sources are able to generate a larger density of source photons and employ very high-quality resonators, such as SRF cavities.

It should be noted that thin barriers can still provide effective EM shielding.
In the radio frequency (RF) regime, $\om\sim 1 \ \text{GHz} \sim 10 \ \mu \text{eV}$ and the length scale $\om^{-1}$ is of order $10 \cm$.
This is much larger than the penetration depth of RF fields into (super)conductors, which can be as small as $\sim 50 \ \text{nm}$, allowing for $d \ll \om^{-1}$.
Further, it is crucial to note that in an evanescent LSthinW setup, it is not required that the photon detector be located within a Compton wavelength $\mA^{-1}$ of the barrier.
Although the virtual hidden photons are restricted to this small region, they act as a localized source of on-shell photons which readily propagate to a distant detector.
It follows that, apart from the thickness of the barrier, there is no exponential suppression relative to any other experimental length scale,
such as the size or frequency of a detection cavity.
Similar insights were noted previously in Ref.~\cite{Hoogeveen:1992nq} within the context of LSW searches for electromagnetically-coupled axions.

In this work, we propose a simple LSthinW setup to search for hidden photons with mass $\mA \gg 1 \ \mu \text{eV}$ using high-quality SRF cavities. As shown schematically in \Fig{fig:LSthinWcartoon}, a driven ``emitter'' cavity sources both SM and hidden fields at frequency $\om \sim 1 \ \text{GHz}$. A narrow superconducting barrier separates and shields the emitter from a quiet ``receiver'' cavity.
The SM field is exponentially attenuated through the barrier over the London penetration depth $\lambda_\tx{L} \sim 50 \ \text{nm}$, while the hidden field is attenuated over $1/\mA \sim 1 \ \text{mm} \times (\text{meV} / \mA)$.
Thus, for a barrier of thickness $10 \ \mu \text{m} \lesssim d \lesssim \mA^{-1}$, the receiver cavity is shielded from the large driven fields, while the evanescent hidden photon field can penetrate into the receiver cavity and excite its resonant modes.
As we show, such an experiment can provide leading sensitivity to hidden photons in the $10 \ \mu \text{eV} - 100 \ \text{meV}$ mass range, whether or not they constitute the dark matter.

The rest of this work proceeds as follows.
In \Sec{sec:hp-lsw}, we introduce the formalism needed to calculate LSthinW signals.
In \Sec{sec:eff-lsw}, we specialize to the evanescent regime and highlight the signal parametrics particular to this case.
We apply this formalism to optimize a simple LSthinW setup in \Sec{sec:eff-srf} and then estimate the sensitivity of several concrete searches.
We conclude in \Sec{sec:dis}.


\section{General Formalism}
\la{sec:hp-lsw}

Here we review the classical equations of motion governing %
SM and hidden photon fields, the production of hidden photons, and their excitation of EM cavities.
We use some of the formalism of Ref.~\cite{Graham:2014sha}, which studied on-shell hidden photons in the far-field limit, but we reformulate some of the key results to highlight the physics of the off-shell evanescent case.
The results of this section are general, however, and can be applied to hidden photons of any mass (provided that $\eps \ll 1$) and are equivalent to the results of Ref.~\cite{Graham:2014sha}.
Later in \Sec{sec:eff-lsw}, we will apply these results to the evanescent case.


\subsection{Classical Field Equations}

We begin by obtaining the classical wave equations for the photon and hidden photon fields in a field basis which diagonalizes the kinetic and mass terms of \Eq{eq:L-kineticmix}.
\Eq{eq:L-kineticmix} is converted by means of the field-redefinitions
\subeqa{eq:to-prop-diag}{
A^\mu &\to A^\mu + \frac{\eps}{\sqrt{1-\eps^2}} \, A^{\prime \, \mu}
\\
A^{\prime \, \mu} &\to \frac{1}{\sqrt{1-\eps^2}} \, A^{\prime \, \mu}
~,
}
which yields
\begin{align}
\la{eq:L-diag}
\Lag &= - \frac{1}{4} \, F_{\mu \nu} F^{\mu \nu} - \frac{1}{4} \, F_{\mu \nu}^\prime F^{\prime \, \mu \nu} + \frac{1}{2} \, \frac{\mA^2}{1 - \eps^2} \, A_\mu^\prime A^{\prime \, \mu}
\nl
&- j_\mu \, \Big( A^\mu + \frac{\eps}{\sqrt{1 - \eps^2}} \, A^{\prime \, \mu}\Big)
~.
\end{align}
In this basis, SM charges couple to both the massless and massive fields, which are themselves uncoupled.
Hence, the wave equations follow immediately from \Eq{eq:L-diag} as
\subeqa{eq:AwaveDiag}{
\la{eq:Awave-unprime}
 \Big( \partial_t^2 - \nabla^2 \Big) \, \vec{A} &= \vec{j} \\
\la{eq:Awave-prime}
 \Big( \partial_t^2 - \nabla^2 + \frac{\mA^2}{1 - \eps^2}  \Big) \, \vec{A}^\p
  &= \frac{\eps}{\sqrt{1 - \eps^2}} \, \vec{j}
~,
}
where $\vec{A}$ and $\vAp$ are the SM and hidden vector potentials, respectively, and $\vec{j}$ is the SM current density.
\Eq{eq:Awave-unprime} is Maxwell's wave equation in Lorenz gauge, with the scalar potential $\phi$ determined by $\partial_t \phi = - \grad \cdot \vec{A} \, $.
The hidden potentials obey an identical condition,\footnote{For massive fields, this is not a choice of gauge, but instead follows from conservation of charge, i.e., $\mA^2 \partial_\mu A^{\prime \, \mu} = - \partial_\mu j^{\prime \, \mu}$.} and the electric and magnetic fields are determined from the potentials in the usual manner for both the SM and hidden fields.

As discussed in Ref.~\cite{Graham:2014sha}, to compute field propagation across a conducting barrier it is useful to define ``visible'' and ``invisible'' linear combinations of fields that couple to or are completely sequestered from SM sources, respectively.
In such a basis, conducting boundary conditions are simple to enforce as they apply only to the visible field and do so in the standard way.
The required transformation can be read off of \Eq{eq:L-diag}.
We define
\subeqa{eq:EvisEinv}{
\la{eq:Evis}
A_\text{vis}^\mu &= \sqrt{ 1-\eps^2 } \, A^\mu + \eps \, A^{\prime \, \mu} \\
\la{eq:Einv}
A_\text{inv}^\mu &= - \eps \, A^\mu + \sqrt{1 - \eps^2} \, A^{\prime \, \mu}
~,
}
and also take $e \to \sqrt{1-\eps^2} \, e$ such that the visible field couples to SM charge via the empirical EM coupling constant.
\Eq{eq:AwaveDiag} becomes
\subeqa{eq:AinvvisFull}{
\la{eq:Avis-full}
\Big(\partial_t^2 - \nabla^2 \Big) \, \vec{A}_\vis
   &= \vec{j} - \frac{\eps \, \mA^2}{1-\eps^2} \, \vec{A}^{\p} \\
\la{eq:Ainv-full}
\Big( \partial_t^2 - \nabla^2 + \frac{\mA^2}{1-\eps^2} \Big) \, \vec{A}_\inv
  &= - \frac{\eps \, \mA^2}{1-\eps^2} \, \vec{A}
~.
}
Note that we have kept the right-hand-side of these equations written in terms of the mass-basis fields $\vec{A}$ and $\vec{A}^\prime$ for simplicity, but since these are linear combinations of $\vec{A}_\vis$ and $\vec{A}_\inv$, \Eq{eq:AinvvisFull} contains only two undetermined fields.
On the right-hand-side of \Eq{eq:Avis-full}, $\vec{A}^\p$ enters analogously to a current that sources $\vec{A}_\vis$. Hence, to leading order in $\eps \ll 1$, we define this \emph{effective current} as
\be
\label{eq:jeffdef1}
\vec{j}_\text{eff} = - \eps \, \mA^2 \, \vec{A}^\p
~.
\ee
This form of the effective current is equivalent to that presented in Ref.~\cite{Graham:2014sha}, $\partial_t \, \vjeff = \eps \, \mA^2 ( \mA^2 \vec{E}^{\p} - \grad \, \grad \cdot \vec{E}^{\p} )$.
This follows from the definition of $\vec{E}^{\p}$ in terms of its potentials, as well as Gauss's law for the hidden field in vacuum $\grad \cdot \vec{E}^{\p} = - \mA^2 \phi^{\p}$, as derived from \Eq{eq:L-diag}.


\subsection{Sourcing the Hidden Field}

Consider an emitter cavity of volume $V_\tx{em}$ which contains a driven, monochromatic SM field at frequency $\om$. 
We want to determine the hidden fields that are produced outside of this cavity.
We work here with the mass-basis fields of \Eq{eq:AwaveDiag}, as this will prove to yield a result which is particularly useful for the evanescent case.

At $\order{\eps^0}$, the emitter cavity is described by the driven massless fields $\vec{E}_\tx{em}$ and $\vec{B}_\tx{em}$ oscillating at frequency $\om$, which vanish within the conducting walls of the cavity and obey conductor boundary conditions\footnote{
$\vec{B}_\tx{em}$ is tangential to the surface and $\vec{E}_\tx{em}$ is normal to the surface.}
on the inner cavity surface $\partial V_\tx{em}$.
These boundary conditions are maintained by charges and currents on $\partial V_\tx{em}$, e.g., the tangential magnetic field is supported by a surface current $\vec{K}_\text{em}$ given by 
\eqa{
    \vec{K}_\tx{em} = - \hat{n} \times \vec{B}_\tx{em}
    ~,
}
where $\hat{n}$ is the unit normal orientated outward to $\partial V_\tx{em}$ and $\vec{B}_\tx{em}$ is evaluated on $\partial V_\tx{em}$.
From \Eq{eq:Awave-prime}, this current sources an $\vAp$  which to leading order in $\eps$ is
\eqa{
\la{eq:Ap-greens}
    \vAp \simeq \frac{\eps}{4 \pi} \, e^{i \om t}
    \int_{\partial V_\text{em}} \hspace{-0.2cm} %
  \diffp{2}{x'}
~ \frac{e^{-i k |\vec{x} - \vxp|}}{|\vec{x} - \vxp|} ~ \vec{K}_\text{em} (\vxp)
~,
}
where the integration is over the inner surface of the emitter cavity and we have defined the wavenumber
\be
k =
\begin{cases}
\sqrt{\om^2 - \mA^2} & (\mA < \om)
\\
-i \, \sqrt{\mA^2 - \om^2} & (\mA > \om)
~.
\end{cases}
\ee

In \Eq{eq:Ap-greens}, we have ignored boundary conditions on the hidden field (i.e., that $\vEvis$ and $\vBvis$ must obey conductor boundary conditions at the surface of the emitter and reciever cavities).
Enforcing these boundary conditions amounts to incorporating small screening corrections to the current in the integrand of \Eq{eq:Ap-greens}, including those generated in the receiver cavity.
However, this only corrects $\vAp$ at $\order{\eps^3}$ and may be ignored.
To see this, note that outside of the emitter cavity, \Eq{eq:Ap-greens} implies that $E_\tx{vis} \sim \eps E^{\p} \sim \order{\eps^2 \, K_\tx{em}}$.
Thus, enforcing boundary conditions perturbs the surface currents by $\order{\eps^2}$.
Therefore, to leading order in $\eps \ll 1$, \Eq{eq:Ap-greens} is valid everywhere outside of the emitter cavity, including in the interior of the receiver cavity.


\subsection{Cavity Excitation via Hidden Photons} 

From \Eq{eq:Avis-full}, we may derive the wave equation for visible electric fields. To leading order in $\eps$ this is
\be
\la{eq:curlcurl}
\grad \times \grad \times \vec{E}_\text{vis} + \partial_t^2 \vec{E}_\text{vis}
\simeq - \partial_t \big( \vec{j} + \vec{j}_\tx{eff} \big)
~.
\ee
We can identify $\vec{E}_\vis$ with the SM electric field.
It obeys the SM wave equation and conductor boundary conditions, except for the introduction of a new source term, which from Eqs.~(\ref{eq:jeffdef1}) and (\ref{eq:Ap-greens}) is
\be
\la{eq:jeffAlt}
  \vjeff
  \simeq - \frac{\eps^2 \, \mA^2}{4 \pi} \, e^{i \om t}
  \int_{\partial V_\text{em}} \hspace{-0.2cm} %
   \diffp{2}{x'}
~  \frac{e^{-i k |\vec{x} - \vxp|}}{|\vec{x} - \vxp|} ~ \vec{K}_\text{em} (\vxp)
~.
\ee

The effective current $\vjeff$ sources visible fields just as SM currents do.
Suppose that $\vjeff$ is monochromatic with frequency $\om$ near the $p^\tx{th}$ mode of the receiver cavity, $\om \simeq \om_p$.
We decompose the visible field in the receiver cavity in terms of this resonant mode as $\vEvis (\vec{x}, t) = \mathcal{E}_p (t) \, \vec{E}_p (\vec{x})$, where $\mathcal{E}_p (t)$ is a complex number which gives the mode's excitation amplitude and phase at time $t$, and the mode profile $\vec{E}_p (\vec{x})$ satisfies
\be
\la{eq:modes1}
\grad \cdot \vec{E}_p =  (\nabla^2 + \om_p^2) \, \vec{E}_p = \vec{E}_p \times \hat{n} \Big|_{\partial V_\text{rec}} \hspace{-0.25cm} = 0
~.
\ee
Here, $\om_p$ is the mode's resonant frequency and $\hat{n}$ is the unit normal to the surface $\partial V_\text{rec}$ of the receiver cavity.
Performing this decomposition in \Eq{eq:curlcurl} yields an equation for the time-evolution of $\mathcal{E}_p(t)$,%
\be
\la{eq:ringup1}
\Big( \partial_t^2 + \frac{\om_p}{Q} \, \partial_t + \om_p^2 \Big) \, \mathcal{E}_p \simeq - \, \frac{\int_\text{rec} \diffp{3}{x} \vec{E}_p^* \cdot \partial_t (\vec{j} + \vjeff)}{\int_\text{rec} \diffp{3}{x} |\vec{E}_p|^2}
~,
\ee
where on the left-hand-side we have inserted a damping term, quantified by the quality factor $Q$ of the receiver resonant mode.
The integrals on the right-hand-side are performed over the volume of the receiver.
Assuming that there are no SM currents within the inner volume of the receiver cavity, $\vec{j}$ can be dropped from the right-hand-side of \Eq{eq:ringup1} since the conducting boundary condition implies that current on the inner walls of the cavity is orthogonal to $\vec{E}_p$ at the boundary.
On resonance ($\om = \om_p$), the visible field in the receiver is
\be
\vEvis (\vec{x}) \simeq  - \, \frac{Q}{\om_p} \, \vec{E}_p (\vec{x}) \, \frac{\int_\text{rec} \diffp{3}{x'} \vec{E}_p^* (\vxp) \cdot \vjeff (\vxp)}{\int_\text{rec} \diffp{3}{x'} |\vec{E}_p (\vxp)|^2}
~,
\ee
which corresponds to a signal power of
\be
\label{eq:Psig1}
P_\text{sig} = \frac{\om_p}{Q} \int_\text{em} \hspace{-0.25cm} \diffp{3}{x} |\vEvis|^2
\simeq \frac{Q}{\om_p} \, \frac{\big| \int_\text{rec}  \diffp{3}{x} \vec{E}_p^* \cdot \vjeff \big|^2}{\int_\text{rec} \diffp{3}{x} |\vec{E}_p|^2}
~ .
\ee


\section{Evanescent Signals}
\la{sec:eff-lsw}

\begin{figure}[t]
\includegraphics[width=.6 \columnwidth]{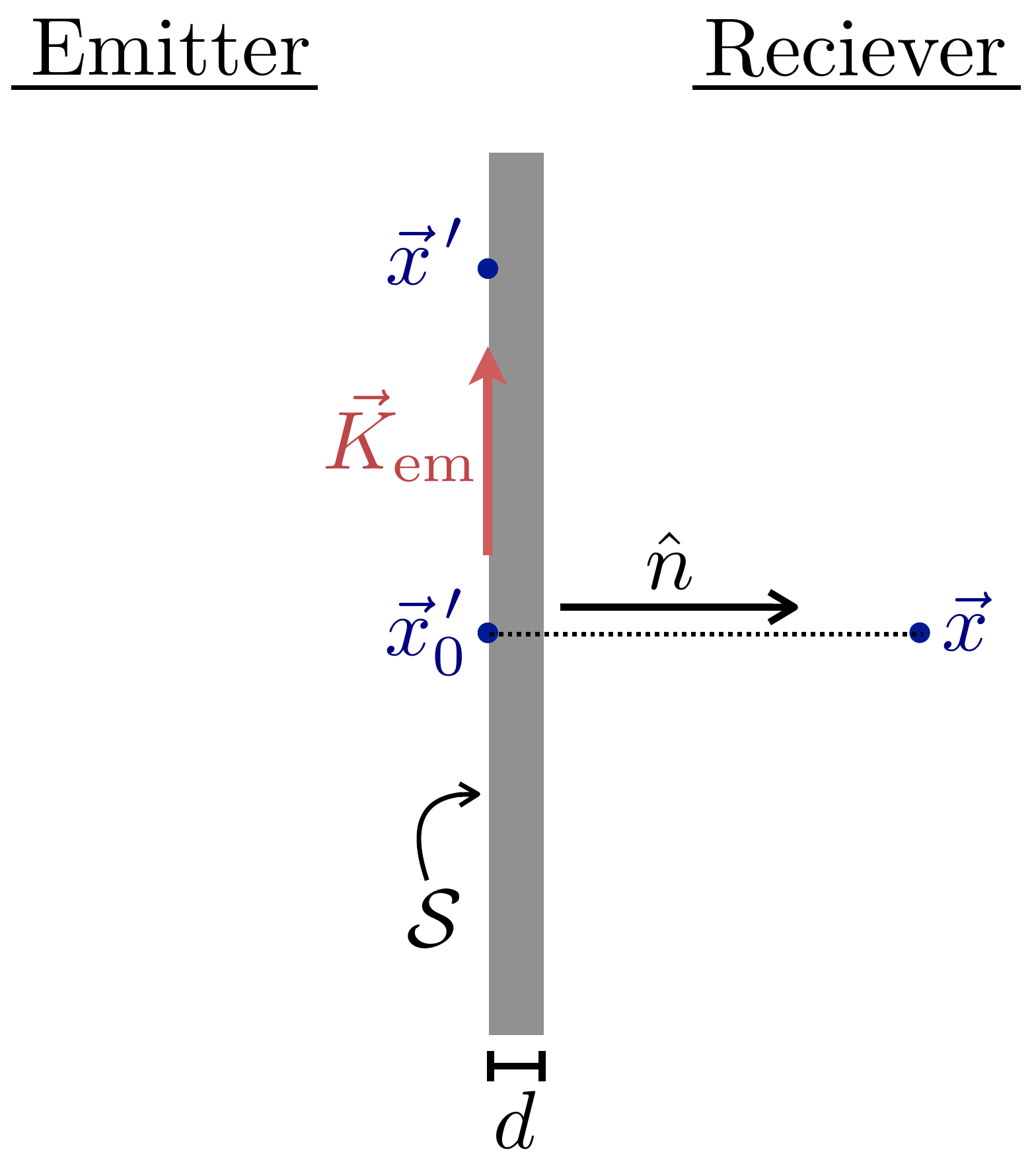}
\caption{A closeup of the thin wall $\mathcal{S}$ separating the emitter and receiver cavities.
To evaluate $\vjeff$ at the point $\vec{x}$ in the receiver cavity, we use the fact that the integral over $\mathcal{S}$ in \Eq{eq:jeffAlt} is heavily weighted near the point $\vxp_0$ in the $\mA \gg \om$ limit.
This yields \Eq{eq:jeffapprox}.
}
\label{fig:calcfig}
\end{figure}

\Eqs{eq:jeffAlt}{eq:Psig1} can be numerically evaluated for any choice of experimental and model parameters to determine the corresponding signal strength.
However, it is useful to also have analytical expressions for simple geometries.
This was done in Ref.~\cite{Graham:2014sha} in the limit that the separation $d$ between the emitter and receiver cavities is much greater than the size of either cavity.
In this work we are interested in the opposite limit, where the cavities are very closely spaced and in the evanescent regime.
As we show below, it is also possible to analytically evaluate $\vjeff$ and $P_\tx{sig}$ in this limit.

To begin, let us take a simple LSthinW setup consisting of two cavity volumes obtained from a single larger volume partitioned by a thin conducting surface $\mathcal{S}$ (as shown in \Fig{fig:calcfig}), whose thickness $d$ is much smaller than both the cavity length and the Compton wavelength of the hidden photon, but much thicker than the EM penetration depth of the partition material.
Consider $\vjeff(\vec{x})$ at a point $\vec{x}$ fixed in the receiver cavity, within a distance $\mA^{-1}$ of $\mathcal{S}$ and a distance much greater than $\mA^{-1}$  from the external walls of the receiver cavity (i.e., the top and bottom of \Fig{fig:LSthinWcartoon}).
The integrand in the expression for $j_\text{eff}$ in \Eq{eq:jeffAlt} is exponentially suppressed for $|\vxp - \vec{x}| \gg 1/\mA$, where $\vxp$ is a point on $\mathcal{S}$.
Thus the $\vxp$ integral is heavily weighted near $\vxp_0$, defined as the point on $\mathcal{S}$ that is closest to $\vec{x}$.
These coordinates are shown in \Fig{fig:calcfig}.
We thus expand $\vec{K}_\text{em}(\vxp)$ around $\vxp_0$ in the integrand of \Eq{eq:jeffAlt}, which is a good approximation for $\vec{x}$ away from the external edges of the receiver cavity and provided that $\om / \mA \ll 1$.
Then, taking $\mathcal{S}$ to be approximately planar near $\vxp_0$, we may extend the integration region to an infinite plane and evaluate  \Eq{eq:jeffAlt} to find 
\begin{align}
\la{eq:jeffapprox}
\vjeff (\vec{x}) \simeq \, & \frac{1}{2} \, \eps^2 \, \mA \, e^{i \om t} \,
e^{- \mA  |\vec{x} - \vxp_0|} ~ \hat{n}  \times \vec{B}_\text{em} (\vxp_0)
\end{align}
up to $\order{\om/\mA}$, where as in \Fig{fig:calcfig}
the $\hat{n}$-axis
is defined along the direction normal to $\mathcal{S}$ pointing from $\vxp_0$ to  $\vec{x}$. %
For later convenience, we factorize the spatial dependence of $\vjeff$ into a dimensionless function $\jhat$ defined as,
\be
\label{eq:jhat}
\vjeff (\vec{x}) \equiv \eps^2 \, \mA \, e^{i \om t} \, e^{- \mA d} \, \bar{B}_\text{em} \,  \jhat (\vec{x})
~,
\ee
where $\bar{B}_\text{em} = \sqrt{\int_\text{em}  \diffp{3}{x} |\vec{B}_\text{em}|^2 / V_\text{em}}$ is the amplitude of the emitter magnetic field RMS-averaged over the emitter cavity volume $V_\text{em}$.

\begin{figure}[!t]
\hspace*{-0.5cm}
\includegraphics[width=1.0\columnwidth]{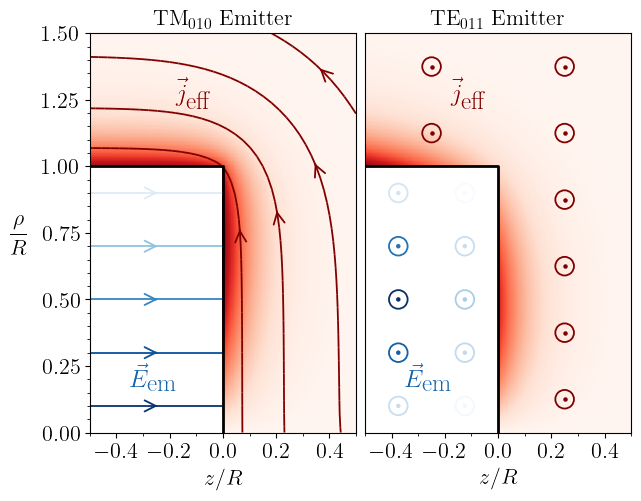}
\caption{The spatial profile (in cylindrical coordinates) of $\vjeff$ outside of a cylindrical emitter cavity of radius $R$ and length $L=R$ for two source modes, $\text{TM}_{010}$ (left panel) and $\text{TE}_{011}$ (right panel).
The hidden photon mass is $\mA = 10 / R$.
Only a portion of the emitter cavity is shown, with its walls indicated by the thick black line.
The emitter electric field profile is shown in \textcolor{mpl_Blues8}{blue}, with darker shading indicating larger field magnitude and arrows indicating direction.
The relative magnitude of the resulting effective current is shown by the \textcolor{mpl_maroon}{red} shading and its direction by the \textcolor{mpl_maroon}{red} arrows.
}
\label{fig:slice}
\end{figure}

\begin{figure}[t]
\hspace*{-1cm}
\includegraphics[width=0.95 \columnwidth]{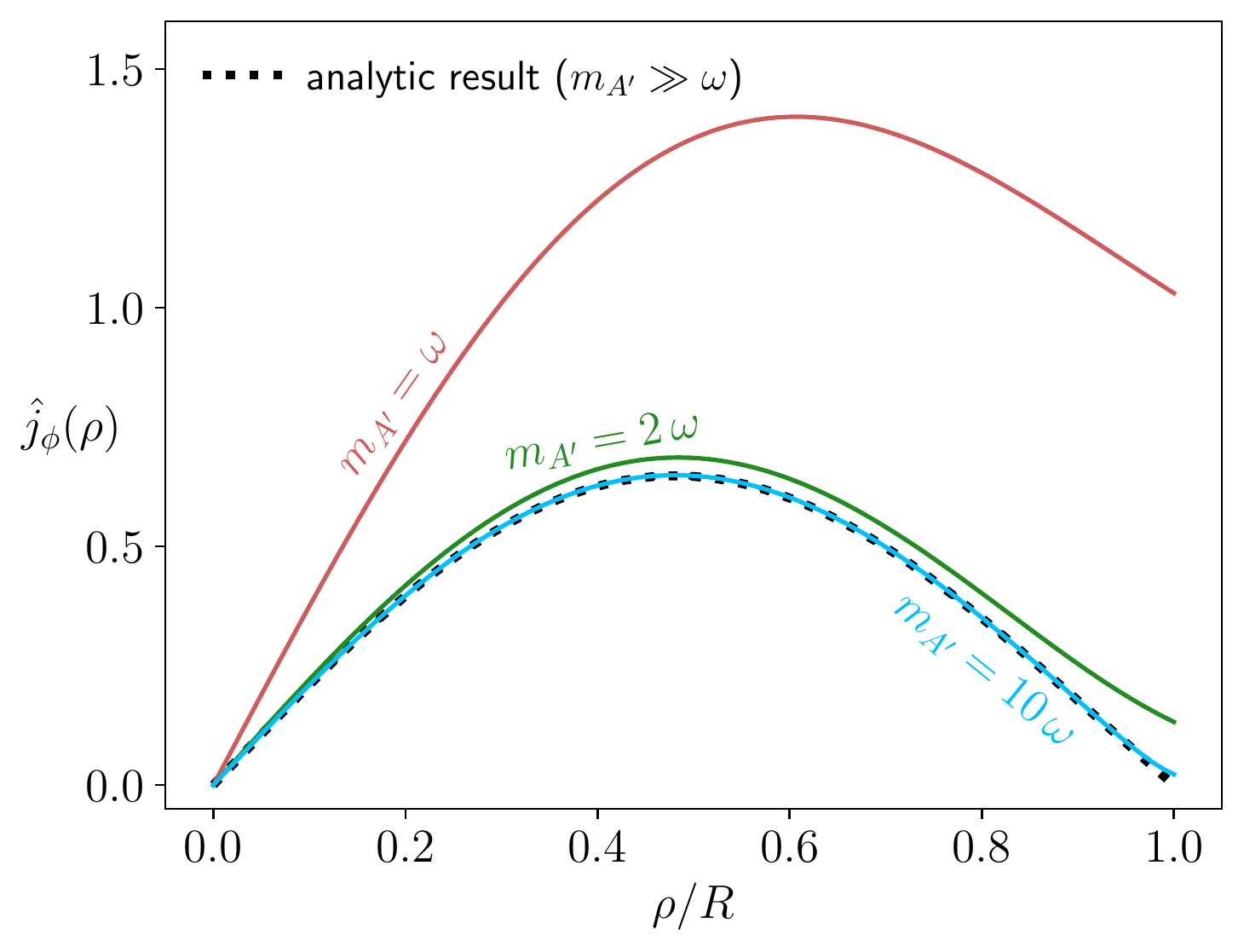}
\caption{Comparison of numerical and analytic evaluations of the $\phi$-component of the effective current profile $\jhat (\vec{x})$ (see \Eq{eq:jhat}) outside the endcap of a cylindrical emitter cavity operating in the $\text{TE}_{011}$ mode.
The solid curves show the numerically-determined radial profiles computed from the integral in \Eq{eq:jeffAlt} (evaluated within a Compton wavelength of the emitter cavity) for various choices of the hidden photon mass $\mA$.
For $\mA \gtrsim \om$, the numerical results agree well with the analytic result of \Eq{eq:jeffapprox}, shown in dotted black.}
\label{fig:jeff}
\end{figure}

An example of $\vjeff$ computed numerically from \Eq{eq:jeffAlt} is shown in \Fig{fig:slice} for two different emitter modes, taking the emitter to be a right-cylindrical cavity.
This displays the features expected from \Eq{eq:jeffapprox} (discussed below), which applies near the emitter wall and away from the side-endcap boundary ($\rho = R$, $z=0$).
In \Fig{fig:jeff}, we compare the analytic result of \Eq{eq:jeffapprox} directly to a numerical evaluation of \Eq{eq:jeffAlt} for a particular emitter field and at various hidden photon masses.
For $\mA \gtrsim 2 \, \om$, $j_\text{eff}$ is well-approximated by \Eq{eq:jeffapprox} to $\lesssim 10 \%$, except within $1/\mA$ of the outer radial edge ($\rho =R$).
In this region, ignoring contributions from surface currents along the radial edge of the emitter cavity is no longer a valid approximation, and the effective current develops a component along the $\hat{z}$ direction.
However, the contribution of such corrections to the signal power is relatively suppressed by the small volume of this region.
In particular, this radial edge contributes to $P_\tx{sig}$ as $\order{\om^4/\mA^4}$, which is sub-dominant for an optimal choice of modes (see \Eq{eq:Psigscaling}).

The effective current in \Eq{eq:jeffapprox} has three key features:
\renewcommand\labelenumi{\theenumi) }
\begin{enumerate}[leftmargin=*]
\setlength{\itemsep}{0.5pt}
\item $j_\text{eff} \propto \mA$.
This follows from \Eq{eq:jeffAlt}, as the relevant length scale in the integral is $1/\mA \ll 1 / \om$.
\item $\vjeff$ is tangential to $\mathcal{S}$.
This is because (away from the external radial edges of the cavity) if $\mA \gg \om$ then the only SM current within a Compton wavelength of the receiver cavity is that running along $\mathcal{S}$.
\item $j_\text{eff} (\vec{x})$ only has weight within a distance $1/\mA$ of $\mathcal{S}$, as evident by the exponential factor in \Eq{eq:jeffapprox}.
\end{enumerate}
These facts have important implications for the signal power and the selection of modes.
In particular, in the numerator of \Eq{eq:Psig1}, the integral of $\vec{E}_p^* \cdot \vjeff$ over the receiver cavity volume involves only the components of the receiver mode electric field that are tangential to $\mathcal{S}$ and within a distance $1/\mA$ from $\mathcal{S}$.
These electric field components are suppressed by $\om / \mA \ll 1$ relative to the typical field value, due to the conducting boundary condition on $\mathcal{S}$.
Taken together, this implies that the overlap of $\jhat$ with the receiver cavity mode scales as  $\int_\text{rec} \hspace{-0.05cm}  \diffp{3}{x'} \vec{E}_p^* (\vxp) \cdot \jhat (\vxp) \propto 1 / \mA^2$. We thus choose to define
a dimensionless overlap parameter $\eta$,
\be
\label{eq:eta}
\eta \equiv \frac{\mA^2}{\om_p^{1/2}} \, \frac{\big| \int_\text{rec} \diffp{3}{x'} \vec{E}_p^* (\vxp) \cdot \jhat (\vxp) \big|}{\sqrt{\int_\text{rec} \diffp{3}{x'} |\vec{E}_p (\vxp)|^2}}
~,
\ee
such that in the evanescent limit and for optimal mode choices, $\eta$ is independent of $\mA$ and set only by cavity and mode geometry.
For an optimal configuration, $\eta \gtrsim \order{1}$ as discussed in
\Sec{sec:modes}.
Using this in \Eq{eq:Psig1}, the signal power reduces to
\be
\label{eq:Psigscaling}
P_\text{sig} \simeq Q \, \bar{B}_\text{em}^2 \, \frac{\eps^4 \, \eta^2}{\mA^2} \, e^{- 2 \mA d}
~.
\ee


\section{LS\texorpdfstring{\MakeLowercase{thin}}{thin}W Design}
\la{sec:eff-srf}

A full design study is beyond the scope of this work.
Here we highlight the key requirements, focusing on those unique to the evanescent regime.
Some practical challenges, such as matching the frequencies of the cavities and improving quality factors, are shared with ongoing on-shell LSW searches~\cite{Romanenko:2023irv}.
A future LSthinW search can make use of their improvements.

The most important design consideration is the use of a thin barrier.
This is optimally as thin as possible to increase the upper mass reach, while still sufficiently thick to provide adequate shielding and to preserve the large field gradients and quality factors associated with SRF cavities.
Below, we also consider the optimal choice of emitter and detector modes as well as cavity shape, as this has important qualitative differences from the on-shell case.
Finally, we give some example experimental parameters and discuss the resulting sensitivity to hidden photons in the  $10^{-6} \ \text{eV} - 10^{-1} \ \text{eV}$ mass range.


\subsection{Thin Barrier}

As discussed above, the largest mass that an LSthinW experiment is sensitive to is dictated by the thickness $d$ of the barrier separating the two cavity volumes, $\mA \sim 1/d$, which is evident by the exponential suppression of $P_\text{sig}$ in \Eq{eq:Psigscaling}.
Although thinner barriers enhance the signal for large masses, the penetration of SM EM fields into superconductors means that a minimum thickness is required to suppress noise from the strong driven fields of the emitter cavity leaking into the detection region.
We conservatively estimate this minimum thickness by demanding that such leakage fields $B_\text{leak} \sim e^{-d/\lambda_\tx{L}} \bar{B}_\text{em}$  be no larger than the signal field $B_\text{sig} \sim Q \, \eps^2 \bar{B}_\text{em} (\om / \mA)$, where $\lambda_\tx{L} \sim 40 \ \text{nm}$ is the London penetration depth of niobium.
This implies a minimum barrier thickness of $d \sim \text{2} \ \mu \text{m}$ for the weakest signals that we consider in this work.
In our reach estimates, we adopt an even more conservative minimum of $d > 10 \ \micron$, corresponding to a maximum hidden photon mass of $\mA \sim 0.02 \ \text{eV}$.

The manufacture and use of a $d \sim 10 \ \micron$ barrier poses additional challenges.
One option would be to utilize $\sim 1 \ \mu \text{m}$ commercial niobium foils~\cite{goodfellow}.
However, maintaining high-$Q$ requires post-fabrication treatments to rid the niobium of material contaminants, and this is not simple to do for very thin surfaces, as it often requires chemical or electropolishing etching treatments that remove the outer $\sim 100 \ \micron$ of material~\cite{Antoine:2014ksx}.
The incorporation of such standalone niobium barriers would be additionally complicated by the fact that they must resist stress induced by vacuum and EM pressure gradients.

An alternative strategy would be to fabricate a rigid barrier by sputtering a few microns of niobium onto a low-loss, insulating substrate, such as sapphire~\cite{sputter}.
In this case, the high-$Q$ of the receiver cavity could be maintained by orientating the exposed sapphire towards the emitter volume and the thin niobium towards the receiver volume.
For such an arrangement, achieving large driven fields in the emitter cavity in the presence of higher loss ($Q \sim 10^9$) sapphire~\cite{braginsky1987experimental,tobar1998anisotropic,krupka1999use,krupka1999complex} requires a larger amount of power to be driven and dissipated in the emitter cavity.
However, for the largest field strengths and volumes that we consider (see Table~\Ref{tab:param}) this corresponds to $\lesssim 100 \ \text{W}$, which can be readily supplied and then dissipated through the liquid helium already required to cool the emitter cavity~\cite{SRFcooling,Hansen_2022}.

Thermal transport through the thin barrier is another concern.
Heat transmission along a standalone $d \sim 10 \ \micron$ niobium barrier or through a dielectric substrate is restricted, causing the temperature at the center of the barrier to be larger than that of the cavity's external walls which are cooled by liquid helium.
If this temperature peak is too large it may quench the superconductivity of the barrier.
This may be mitigated, for instance, by incorporating helium cooling channels through the substrate itself, but we leave a dedicated study of such designs to future work.
To account for this, we consider two representative barrier thicknesses: $d=0.5 \ \tx{mm}$ (which is similar to the wall thickness of existing SRF cavities and for which the above challenges are not expected to be a concern~\cite{Singer:2014kbq}) and $d=10\ \micron$.


\subsection{Cavity and Mode Geometry}
\la{sec:modes}

We consider a simple arrangement of two coaxial right-cylinder cavities of radius $R$ and length $L$ formed by partitioning a single cylinder of radius $R$ with a planar barrier parallel to its endcaps.
From the structure of the effective current in \Eq{eq:jeffapprox}, we can ascertain that the optimal receiver mode has electric field components tangential to the barrier surface.
This is the case for the $\text{TE}_{011}$
mode, which is azimuthal, $\vec{E} (\vec{x}) \propto \, J_1(\alpha_{11} \rho / R) \, \sin{(\pi z / L)} \, \hat{\phi}$, where the radial coordinate $\rho$ spans $[0,R]$, $z$ spans $[0,L]$, and $\alpha_{11} \simeq 3.83$ is the first zero of $J_1$.
\Eq{eq:jeffapprox} along with $\hat{n} = \hat{z}$ implies that to excite this receiver mode, the magnetic field of the emitter should possess radial components near the endcap.
This is also satisfied by the $\text{TE}_{011}$ mode, since $\vec{B} (z=L) \propto \, J_1(\alpha_{11} \rho / R) \, \hat{\rho}$.
Hence, we expect optimal overlap when both cavities are operated in the $\text{TE}_{011}$ mode.
This is also demonstrated schematically by the right panel of \Fig{fig:slice}.
Analytically evaluating the overlap parameter of \Eq{eq:eta} for this mode choice, we find
\eqa{
\la{eq:eta_full}
\lim_{\mA \gg \om} \eta^2
= \frac{\pi^5 R^5}{L^2 \big( \pi^2 R^2 + \alpha_{11}^2 L^2 \big)^{3/2}}
~,
 }
such that $\eta \simeq 1.6$ for $R=L$ and $\mA \gg \om$.

The dimensions of the emitter/receiver cavity also strongly impact the signal power.
In particular, from \Eq{eq:eta_full} we see that the evanescent signal grows significantly with the aspect ratio $R/L$ as
\be
\label{eq:RggL}
\lim_{R \gg L \gg \mA^{-1}} \eta = \pi R / L
~.
\ee
This is expected in the evanescent limit.
Since $j_\text{eff}$ is peaked within one Compton wavelength of the barrier, decreasing the receiver length increases the fraction of the receiver volume
which contains appreciable effective current.
\Eqs{eq:Psigscaling}{eq:RggL} imply
that the signal power is independent of the receiver volume provided that $\mA \gg \om, R^{-1}, L^{-1}$.
We thus consider as an example a search with $R/L=50$.
Note that such a geometry causes a suppression of the on-shell ($\mA < \om$) signal, as evident in \Fig{fig:reach}, due to decreased receiver volume and incoherence of the emitted hidden photon field over the emitter cavity, as now $R \gg 1/\omega \simeq L / \pi$.
Thus, such a geometry is purely optimized for evanescent hidden photons.

While there is a strong overlap for a $\text{TE}_{011}$ emitter to excite a $\text{TE}_{011}$ receiver in the evanescent limit, this does not generally hold for matched cavity modes.
This is evident in the left panel of \Fig{fig:slice}, which shows $\vjeff$ sourced by a $\text{TM}_{010}$ emitter mode.
In the coaxial region ($\rho < R$, $z>0$), $\vjeff$ is primarily radial but the mode has electric fields purely in $\hat{z}$.
Thus, $\vjeff \cdot \vec{E}_p^*$ is non-zero only within $1/\mA$ of the outer radial edge of the endcap ($\rho = R$, $z=0$).
As a result, for both cavities in the $\text{TM}_{010}$ mode, $\eta$ is suppressed by $\order{\om/\mA} \ll 1$ relative to that of the $\tx{TE}_{011}$ case considered above.\footnote{\Fig{fig:slice} implies that large overlap is possible for $\text{TM}_{010}$ using a different cavity arrangement, such as nested concentric cavities.}

It is natural to ask if evanescent hidden photons can be searched for in a setup minimally modified from existing RF LSW efforts, such as Dark SRF~\cite{Romanenko:2023irv}, simply by taking the cavity separation to be much smaller than the $d \sim \omega^{-1} \sim 10 \cm$ gap currently used.
The Dark SRF setup can be well-approximated as two cylindrical cavities operating in the $\text{TM}_{010}$ mode.
This was chosen to target the longitudinal $A^\prime$ polarization, which provides enhanced sensitivity in the on-shell $\mA \ll \om$ regime~\cite{Graham:2014sha}.
However, as discussed above, this mode configuration has a suppressed overlap in the evanescent limit.
Therefore, in addition to decreasing the cavity separation, a change in modes is needed for an effective search for hidden photons heavier than $\sim \text{few} \times \mu \text{eV}$.
It is also interesting to note that the CROWS LSW experiment conducted a search employing the optimal $\text{TE}_{011}$ configuration discussed here~\cite{Betz:2013dza}.
However, their cavity separation $d \simeq \om^{-1}$ was too large to enable enhanced sensitivity to evanescent hidden photons.


\subsection{Hidden Photon Sensitivity}
\la{sec:sensitivity}

The reach of an optimized LSthinW setup is estimated by the signal-to-noise ratio, $\text{SNR} = P_\text{sig} / P_\text{noise}$, where the noise power $P_\text{noise}$ is assumed to be dominated by thermal occupation of the receiver mode.
We take $\tx{SNR} = 5$ to determine the projected sensitivity. 
If the phase of the emitter cavity is not actively monitored, the noise power is given by the Dicke radiometer equation, $P_\text{noise} \simeq T \sqrt{\delta \om /  2 \pi t_\text{int}}\, ,$ where $T$ is the receiver cavity temperature, $\delta \om$ is the analysis bandwidth which we take to be the receiver bandwidth $\om/Q$, and $t_\text{int}$ is the experimental integration time.
Following Ref.~\cite{Graham:2014sha}, an active monitoring of the emitter field's phase allows for enhanced sensitivity, corresponding to an effective thermal noise power of $P_\text{noise} \simeq T / t_\text{int}$.
We will consider both possibilities in our estimates below.

\begin{table}[t!]
\setlength\extrarowheight{3pt}
\setlength{\tabcolsep}{3pt}
\centering
\begin{tabular}{c c c c c c c }
\hline\hline
 LSthinW & $d$ & $R$ & $L$ & $Q$ & $t_\text{int}$ & \text{readout}
 \\ \hline
I & $0.5 \ \text{mm}$ & $10 \ \text{cm}$ & $10 \ \text{cm}$ & $10^{10}$  & $4 \ \text{hr}$ &  \text{power}
\\ 
II & $0.5 \ \text{mm}$ & $10 \ \text{cm}$ & $10 \ \text{cm}$ & $10^{12}$   & $1 \ \text{yr}$ & \text{phase}
\\ 
III & $10 \ \mu \text{m}$ & $15 \ \text{cm}$ & $0.3 \ \text{cm}$ & $10^{12}$  & $1 \ \text{yr}$ & \text{phase}
\\ \hline\hline
\end{tabular}
\caption{Benchmark experimental parameters for the three projections shown in \Fig{fig:reach}.
For each, the peak magnetic field on the emitter cavity wall is fixed to $B=100 \ \tx{mT}$ and the readout temperature is $T = 2 \ \tx{K}$.
Each setup assumes cylindrical cavities with radius $R$, length $L$, and quality factor $Q$.
The cavities are assumed to be aligned along a common axis, separated by a wall of thickness $d$.
$t_\tx{int}$ is the total integration time.
The last column denotes whether the readout involves only a measurement of power in the receiver, or additionally employs an active monitoring of the emitter field's phase.
See \Sec{sec:eff-srf} for details.}
\label{tab:param}
\end{table}

\begin{figure*}[t]
\hspace*{-1.5cm}
\includegraphics[width=1.5 \columnwidth]{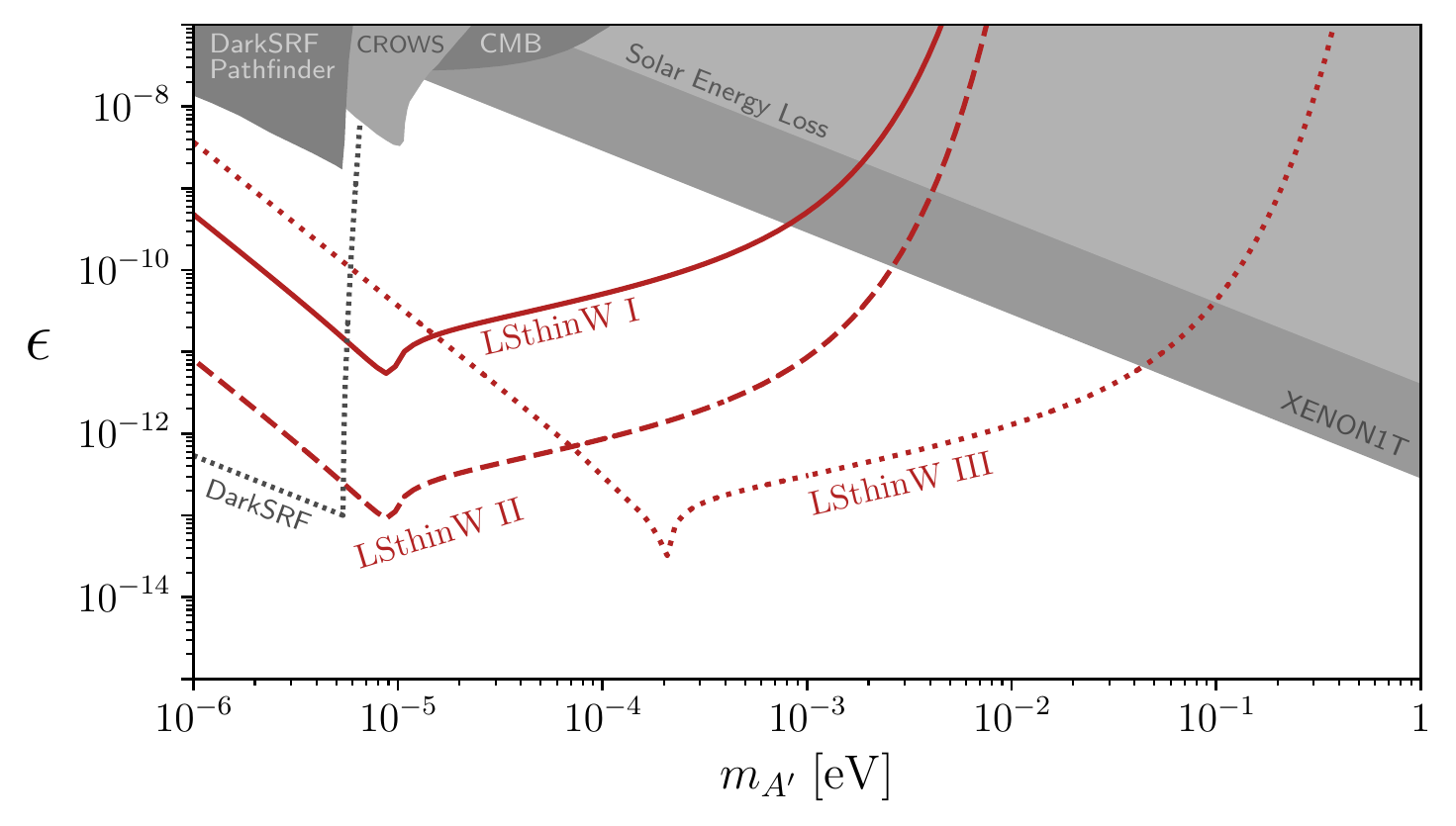}
\caption{
The projected reach of three representative LSthinW setups employing SRF cavities, as shown in \textcolor{mpl_firebrick}{red}.
All the searches assume coaxial right-cylindrical cavities, with the emitter cavity driven in the $\tx{TE}_{011}$ mode and the receiver cavity tuned to have a $\tx{TE}_{011}$ mode of matching frequency.
LSthinW I (solid \textcolor{mpl_firebrick}{red}) assumes readily achievable experimental parameters and employs cavities of equal aspect ratio ($R=L=10 \cm$) separated by a barrier of thickness $d = 0.5 \ \text{mm}$.
LSthinW II (dashed \textcolor{mpl_firebrick}{red}) and III (dotted \textcolor{mpl_firebrick}{red}) use improved experimental parameters, similar to the design goals of the ongoing Dark SRF experiment (whose projected reach is shown in dotted dark \textcolor{gray}{gray}~\cite{Berlin:2022hfx}).
LSthinW II and III are complementary searches, using different cavity geometries to specialize to smaller and larger $\mA$, respectively.
LSthinW II uses the same cavity shape and wall thickness as LSthinW I, whereas LSthinW III employs a pancake-like cavity ($R=15 \cm, \, L=0.3\cm$) and a smaller barrier thickness, $d = 10 \ \micron$.
For more details, see Table \ref{tab:param} and \Sec{sec:sensitivity}.
Also shown in shaded \textcolor{gray}{gray} are existing constraints on  kinetically-mixed hidden photons, taken from the repository Ref.~\cite{AxionLimits}.
These include limits derived from the CROWS LSW experiment~\cite{Betz:2013dza}, the Dark SRF LSW Pathfinder run~\cite{Romanenko:2023irv}, XENON1T~\cite{An:2020bxd}, CMB spectral distortions~\cite{Mirizzi:2009iz,Caputo:2020bdy,Garcia:2020qrp}, and solar energy loss~\cite{An:2013yfc,Redondo:2013lna}.}
\la{fig:reach}
\end{figure*}

Operating at low temperature is advantageous, and existing RF LSW searches optimize their cooling strategy by cooling the receiver much more than the emitter~\cite{Romanenko:2023irv}.
This may not be feasible for an LSthinW search, as the use of a thin barrier with no vacuum gap places the emitter and receiver in thermal contact.
The entire system likely needs to be cooled to the temperature of the receiver cavity, and for this reason we adopt a conservative readout temperature of $T = 2 \ \tx{K}$.

The projected sensitivities of three distinct LSthinW setups are shown in \Fig{fig:reach}, assuming coaxial right-cylindrical cavities operated in the $\text{TE}_{011}$ mode.
Experimental parameters for each are given in Table~\Ref{tab:param}.
In all cases we take $T = 2 \ \tx{K}$ and we set the magnetic field in the emitter to have a peak value along the cavity walls of $100 \ \tx{mT}$, slightly below the critical field of niobium.
For the ``LSthinW I'' setup, we consider cavities of aspect ratio $R/L = 1$, a relatively thick separating barrier $d = 0.5 \ \tx{mm}$, and generally conservative parameters regarding cavity volume, quality factor, and integration time.
For ``LSthinW II,'' we adopt the same cavity and mode geometry, but optimize the other parameters to the design goals of the existing Dark SRF collaboration~\cite{Romanenko:2023irv}.
Our estimates demonstrate that either of these two setups would enable sensitivity to a large range of unexplored parameter space for hidden photons of mass $10^{-5} \ \text{eV} < \mA < 10^{-3} \ \text{eV}$.
``LSthinW III'' employs parameters optimized for larger hidden photon masses, $10^{-4} \ \text{eV} < \mA < 10^{-1} \ \text{eV}$, using cavities with $R/L = 50$ and a thinner barrier of $d = 10 \ \micron$.
Also shown as shaded gray in \Fig{fig:reach} are existing limits on the existence of kinetically-mixed hidden photons.

In \Fig{fig:reach}, the sensitivity is maximized in each setup when the hidden photon mass is equal to the cavity frequency, $\om = \sqrt{(\pi/L)^2 + (\alpha_{11} / R)^2} \, $.
The cavity frequency for LSthinW III is larger than that of LSthinW I and LSthinW II due to the smaller cavity length.
The scaling $\eps \propto \mA^{1/2}$ for $\mA > \om$ follows from \Eq{eq:Psigscaling}.
For $\mA < \om$ we find $\eps \propto \mA^{-2}$, as expected from Ref.~\cite{Graham:2014sha} for an on-shell search in the ``transverse configuration.''
These two power laws suggest that the optimal sensitivity occurs for $\mA \simeq \om$. 
But there is additionally a resonant peak at $\mA = \om$, evident in \Fig{fig:reach}.
At this critical mass, the hidden photon field is sourced with wavenumber $k = 0$, giving maximally constructive interference in the integral of \Eq{eq:Ap-greens}.
The width of this feature is narrower for cavities of larger aspect ratio $R/L \gg 1$, as $R \gg \om^{-1}$ implies a larger phase incoherence over the emitter surface for a small deviation of $\mA$ away from $\om$.


\section{Discussion}
\la{sec:dis}

The highest mass accessible to light-shining-through-wall experiments is dictated not by the frequency of the driven field but by the inverse of the distance separating the emission and detection regions.
Thus, the upper part of the mass range that can be explored can be significantly enlarged in a  ``light-shining-through-\emph{thin}-wall" (LSthinW) experiment.
We have focused on thin superconducting barriers, since their small penetration depth implies that only several microns is needed for sufficient shielding.
In particular, such an experiment employing superconducting cavities operating at $\sim \text{GHz}$ frequencies could have exquisite sensitivity to hidden photons as heavy as $\sim 0.1 \eV$.
The development of thin barriers, as opposed to higher-frequency sources, is a natural alternative to enlarging the mass reach to new particles and has strong synergistic overlap with other efforts, such as future versions of the ARIADNE experiment~\cite{ARIADNE:2017tdd,Fosbinder-Elkins:2017osp}, which will tentatively utilize $\sim 100 \ \mu \text{m}$ niobium shields.

A unique feature of this setup is its ability to produce and detect
particles of mass up to $\sim 0.1 \eV$, regardless of whether they constitute the dark matter of our Universe.
Indeed, the power of this approach is highlighted by the fact that most experiments operating in this mass range are often hindered by the difficulty in operating low-loss resonators and photon detectors at $\text{meV} \sim \text{THz}$ frequencies~\cite{Caldwell:2016dcw,Gelmini:2020kcu,BREAD:2021tpx,Chiles:2021gxk,Chen:2022pyd,Fan:2022uwu}.
In fact, our projections suggest that the reach of an LSthinW experiment may even exceed that of certain dark matter detectors~\cite{Fan:2022uwu}, even though the latter benefit from the assumed presence of a hidden photon dark matter background.

In this work, we have discussed various setups in which an LSthinW experiment using electromagnetic cavities may be realized, but dedicated design efforts are needed to fully bring this proposal to light.
While we have considered simple vacuum cavities with a conducting barrier, other arrangements may be advantageous and deserve further study.
For example, the mass suppression of the signal power in \Eq{eq:Psigscaling} is not fundamental, but results from the vanishing of the receiver cavity mode near the conducting barrier.
It may be that the use of absorbing shield materials or novel mode structures with more support near the barrier can avoid this suppression and generate an improved scaling of $P_\tx{sig} \propto \mA^0$.
Finally, in addition to searching for hidden photons, 
the proposed LSthinW approach may also be applied to enlarge the mass reach for related experiments searching for different types of particles, such as electromagnetically-coupled axions~\cite{Janish:2019dpr,Gao:2020anb} and millicharged particles~\cite{Berlin:2020pey}.
We leave these investigations to future work.


\section*{Acknowledgements}
We would like to thank Paddy Fox, Timergali Khabiboulline, Sam Posen, Paul Riggins, Vladimir Shiltsev, and Slava Yakovlev for valuable conversations.
This material is based upon work supported by the U.S. Department of Energy, Office of Science, National Quantum Information Science Research Centers, Superconducting Quantum Materials and Systems Center (SQMS) under contract number DE-AC02-07CH11359.
Fermilab is operated by the Fermi Research Alliance, LLC under Contract DE-AC02-07CH11359 with the U.S. Department of Energy.

\bibliography{main}

\end{document}